\documentstyle[12pt,aasms4,epsf]{article}

\begin{document}
\setlength{\baselineskip}{12pt}
\def\gapprox{\mathrel{\vcenter{\offinterlineskip \hbox{$>$}
    \kern 0.3ex \hbox{$\sim$}}}}
\def\lapprox{\mathrel{\vcenter{\offinterlineskip \hbox{$<$}
    \kern 0.3ex \hbox{$\sim$}}}}
\def\plotone_reduction#1#2{\centering \leavevmode        %KT
\epsfxsize=#2\columnwidth \epsfbox{#1}}

\title{The Effect of Resistivity on the Nonlinear Stage of
the Magnetorotational Instability in Accretion Disks}

\author{Timothy P. Fleming, James M. Stone\altaffilmark{1}}
\affil{Department of Astronomy, University of Maryland, College Park, MD 20742}

\author{and John F. Hawley}
\affil{Department of Astronomy, University of Virginia, P.O. Box 3818,
University Station, Charlottesville, VA 22903}

\altaffiltext{1}{also Institute of Astronomy, University of Cambridge, 
Madingley Road, Cambridge CB3 0HA, UK}

\begin{abstract}

We present three-dimensional magnetohydrodynamic simulations of the
nonlinear evolution of the magnetorotational instability (MRI) with a
non-zero Ohmic resistivity.  The simulations begin from a homogeneous
(unstratified) density distribution, and use the local shearing-box
approximation.  The evolution of a variety of initial field
configurations and strengths is considered, for several values of
constant coefficient of resistivity $\eta$.  For uniform vertical and
toroidal magnetic fields we find unstable growth consistent with the
linear analyses; finite resistivity reduces growth rates, and, when
large enough, stabilizes the MRI.  Even when unstable modes remain,
resistivity has significant effects on the nonlinear state.  The
properties of the saturated state depend on the initial magnetic field
configuration.  In simulations with an initial uniform vertical field,
the MRI is able to support angular momentum transport even for large
resistivities through the quasi-periodic generation of axisymmetric
radial channel solutions rather than through the maintenance of
anisotropic turbulence.  Reconnective processes rather than parasitic
instabilities mediate the resurgent channel solution in this case.
Simulations with zero net flux show that the angular momentum transport
and the amplitude of magnetic energy after saturation are significantly
reduced by finite resistivity, even at levels where the linear modes
are only slightly affected.  The MRI is unable to sustain angular momentum
transport and turbulent flow against diffusion for $Re_M \lapprox
10^{4}$, where the Reynolds number is defined in terms of the disk
scale height and sound speed, $Re_M = c_s H /\eta$.  As this is close
to the Reynolds numbers expected in low, cool states of dwarf novae,
these results suggest that finite resistivity may account for the low
and high angular momentum transport rates inferred for these systems.

\end{abstract}


\section{Introduction}

In recent years a more complete understanding of the origin of angular
momentum transport and turbulence in accretion disks has emerged.  The
discovery that weak magnetic fields render a differentially rotating
plasma unstable (Balbus \& Hawley 1991; 1992) has directed the focus
of research efforts to magnetohydrodynamic (MHD) processes and the
onset and evolution of the magnetorotational instability (MRI).  The
MRI is a linear, local instability whose existence is independent of
both field orientation and strength.  The action of the MRI directly
produces outward angular momentum transport, as required for accretion
disks to accrete.  

Studies of the nonlinear evolution and saturation of the MRI are
required in any comparison between theory and observation; such studies
rely on numerical MHD simulations.  Many such numerical investigations
have been carried out in recent years.  Local three-dimensional MHD
simulations have shown that turbulence is initiated and sustained by
the MRI (Hawley, Gammie, \& Balbus 1995, hereafter HGB; Brandenburg et
al. 1995; Stone et al. 1996), and that this turbulence supports a
significant outward flux of angular momentum.  Both the turbulent
energy and angular momentum fluxes are dominated by Maxwell rather than
Reynolds stress.  Simulations that begin with a vanishing mean flux
have been used to examine the implications of the MRI for dynamo action
in accretion disks (Brandenburg et al.\ 1995; Hawley, Gammie, \& Balbus
1996, hereafter HGB2) These have demonstrated that the MRI is capable
of amplifying and sustaining an initially weak magnetic field for many
resistive decay times, thus satisfying the minimum definition of a
dynamo.  Moreover, these studies show that this process cannot be
described by kinematic dynamo theory:  the effect of the Lorentz force
on the flow can never be ignored, even when the field is weak.  A
comprehensive review of these and many other results pertaining to the
MRI and angular momentum transport in accretion disks is given by
Balbus \& Hawley (1998, hereafter BH).

Most of the numerical studies of the nonlinear stage of the MRI
reported to date have adopted the assumption of ideal MHD, i.e.
infinite conductivity, so that the field is perfectly frozen-in to the
gas.  However, in cold, dense plasmas such as might be expected at the
centers of protostellar disks (Stone et al. 1998), or disks in dwarf
novae systems (Gammie \& Menou 1997), the ionization fraction may
become so small that this approximation no longer holds.  Obviously,
the appropriate representation of some region of interest within a
partially ionized plasma depends on the specific densities,
temperatures, and ionization fractions therein.  Different nonideal MHD
effects can be important for different physical conditions (see, e.g.,
Parker 1979).  For example, the ambipolar diffusion regime occurs when
the neutral-ion collision time is too long to prevent the gas from
drifting across magnetic field lines; generally this corresponds to low
densities and ionization fractions (more precisely, when the ratio
${\cal R}$ of the product of the ion and electron gyrofrequencies to
the product of the electron-neutral and ion-neutral collision
frequencies is ${\cal R} > 1$).  The linear properties of the MRI in
the ambipolar diffusion limit have been studied in detail by Blaes \&
Balbus (1994).  Their primary conclusion is that the MRI will grow
provided the neutral-ion collision frequency is greater than the
orbital frequency.  The resistive regime occurs when collisions between
the charge carrying species and neutrals damp electric currents (and
therefore magnetic fields) in the plasma, a process equivalent to Ohmic
resistivity.  Resistive effects generally dominate in high density but
weakly ionized plasmas (more precisely, when ${\cal R} < 1$).  The
local linear stability properties of the MRI with simple resistivity
have been examined for both vertical fields (Jin 1996; BH) and toroidal
fields (Papaloizou \& Terquem 1997).  The global stability of resistive
disks has been examined by Sano \& Miyama (1999).  The effect of
resistivity on the linear instability is straightforward:  if Ohmic
diffusion is sufficiently rapid, it can stabilize the MRI.  Recently
Wardle (1999) has explored the linear properties of the MRI in the a
third regime where the conductivity tensor is dominated by the Hall
effect.  In this limit the MRI exhibits interesting new behavior,
including a loss of symmetry with respect to the sign of the background
magnetic field.

Just as the linear properties of the MRI in these regimes
have been investigated, the nonlinear evolution has also been simulated
for a variety of limits.  In the so called ``strong-coupling limit" (in
which the ion inertia is ignored, and the ion density is assumed to be
a simple power law of the neutral density), the effect of ambipolar
diffusion is to add a nonlinear diffusion term to the induction
equation.  As a test of a numerical algorithm to solve this term,
MacLow et al (1995) reported two-dimensional simulations of the initial
growth of the MRI; their results were in agreement with the stability
criterion derived by Blaes \& Balbus (1994).  Brandenburg et
al.\ (1995) performed three-dimensional simulations of the MRI in
stratified disks; some of their models included the effects of
ambipolar diffusion.  They found that sufficiently large diffusivity
could damp the MHD turbulence, consistent with the conclusions of Blaes
\& Balbus (1994).

Hawley \& Stone (1998) carried out a full ion-neutral
simulation of the MRI in weakly ionized plasmas, where the ions and
neutrals are treated as separate fluids coupled only through a
collisional drag term.  Although they found close agreement with the
Blaes and Balbus linear results, the structure and evolution of the
saturated state of the MRI in weakly coupled fluids is more complex.
Full turbulence is produced only when the collision frequency is
$\gapprox 100$ times the orbital frequency.  At lower collision
frequencies, the nonlinear turbulence is increasingly inhibited by the
neutrals, resulting in significantly lowered angular momentum transport
rates.  These results illustrate that it is possible for nonideal
effects to have significant consequences for the nonlinear evolution
even when their impact on the linear instability is slight.

The nonlinear evolution of a differentially rotating flow with a
nonzero resistivity represents a more straightforward simulation
regime.  In their study of dynamo action associated with the MRI, HGB2
presented two simulations in which explicit resistive effects were
included.  These indicated that saturation amplitude and angular
momentum transport rates could be significantly decreased by a large
resistivity.  More recently, Sano, Inutsuka, \& Miyama (1998; hereafter
SIM) explored saturation of the 2D channel solution in the presence of
strong resistivity.  They found reconnection of magnetic field lines
across the channels could act as a saturation mechanism.

In this paper we will explore in greater depth the nonlinear behavior
of the MRI in a single fluid with a finite resistivity.  We present an
extensive series of resistive MHD simulations in three-dimensions.  We
study a variety of initial field configurations and strengths over a
wide range of resistivities.  We find that there are substantial
differences between the nonlinear evolution of the MRI in the presence
of a net flux compared to the evolution at the resistivity with zero
volume-averaged flux.  This behavior arises because Ohmic dissipation
can never destroy a net field, i.e., one which is supported by currents
outside the simulation domain.

As has already been pointed out (BH), a critical dimensionless
parameter which may control the behavior of the MRI in dissipative
disks is the magnetic Prandtl number, i.e., the ratio of the
coefficients of viscosity and resistivity.  The present study does not
include a physical viscous dissipation; some effective viscous
dissipation is, of course, already present due to numerical effects.
Thus, the exploration of the role of magnetic Prandtl number in
determining the nonlinear outcome of the MRI must await future
studies.

This paper is organized as follows.  In \S2, we discuss our numerical
method (including the extension of the ZEUS algorithm to model highly
resistive plasmas), and initial and boundary conditions that
characterize the simulations.  In \S3, we discuss the results of
simulations with uniform vertical fields, vertical fields with zero net
flux, and uniform azimuthal fields.  Our conclusions are presented in
\S4.

\section{Method}

\subsection{Equations and Algorithms}

Our computational model is based on the shearing box approximation
developed by HGB.  This approximation uses a local expansion of the
equations of motion about a fiducial point of radius $R_\circ$ in
cylindrical coordinates $(R, \phi ,z)$.  By considering a region whose
extent is much less than $R_\circ$, one can define a local set of
Cartesian coordinates $x=R-R_\circ, y=R_\circ( \phi - \Omega t), z=z$
that corotate with the disk.  Using $|x|/R_\circ \ll 1$, we expand the
equations of motion to first order in $|x|/R_\circ$  to obtain the
local equations of compressible MHD (HGB):
\begin{equation}
{{\partial \rho} \over {\partial t}} + {\bf \nabla} \cdot (\rho {\bf v}) = 0
\label{continuity}
\end{equation}
\begin{equation}
{{\partial {\bf v}} \over {\partial t }} + {\bf v} \cdot \nabla {\bf v} 
= -{1 \over 
\rho} \nabla \left (P + {B^2 \over 8\pi} \right) + {{{\bf B} \cdot
\nabla {\bf B}} \over 4\pi \rho} - 2 {\bf \Omega} \times {\bf v} 
+ 3 \Omega^2 x {\bf {\hat{x}}}
\label{momentum}
\end{equation}
\begin{equation}
{\partial {\bf B} \over {\partial t }} = {\bf \nabla} \times [({\bf v} \times 
{\bf B}) - \eta {\bf J}],
\label{induction}
\end{equation}
\begin{equation}
{\partial {\rho \epsilon} \over {\partial t}} + {\bf \nabla} \cdot (\rho 
\epsilon {\bf v}) + P({\bf \nabla} \cdot {\bf v}) - \eta \bf J^2 = 0
\label {energy}
\end{equation}
where $\epsilon$ is the specific internal energy, ${\bf J} = {\bf
\nabla} \times {\bf B}$ is the current density, and the other symbols
have their usual meaning.  For our study, the magnetic diffusivity
$\eta$ is spatially uniform and time independent.  The $z$ component of
gravity is ignored; consequently, there are no vertical buoyancy
effects.  We adopt an adiabatic equation of state \begin {equation} P =
\rho \epsilon (\gamma - 1) \label {1e} \label{eos} \end{equation} with
$\gamma= 5/3$.  A hydrodynamic equilibrium solution to equations
(1)--(4) is constant density and pressure and
a uniform shear flow, ${\bf v}=-(3/2) \Omega x {\bf {\hat{y}}}$.

The shearing box approximation employs strictly periodic boundary
conditions in the angular ($y$) and vertical ($z$) directions, and
shearing-periodic boundary conditions in the radial ($x$) direction.
These boundary conditions and their implementation are described in
more detail in HGB.  Briefly, faces along the $x$ directions are
periodic initially  but subsequently shear with respect to each other.
Any fluid element that travels off the outer radial boundary reappears
at the lower radial boundary at the corresponding sheared position.

The above equations of MHD are solved using the ZEUS code (Stone and
Norman 1992a; 1992b).  ZEUS is a time explicit MHD code based on finite
differences that uses the Method of Characteristics -- Constrained
Transport (MOCCT) algorithm (Hawley \& Stone 1995) to evolve both the
induction equation and the Lorentz force.  The advantage of MOCCT is
that it evolves the magnetic field in such a way as to maintain the
constraint $\nabla \cdot \bf{B} = 0$.  Key to this property is the use
of Stoke's Law to write the induction equation in integral form: the
rate of change of the magnetic flux through any face of a computational
zone is then simply the line integral of the electromotive force (emf)
around the edges of the face (Evans \& Hawley 1988).  To extend the
method to include resistivity, we use an operator split solution
procedure in which the MOCCT technique is used to update the first term
on the RHS of (3) and then the constrained transport
formalism is again used to update the magnetic flux using an effective
emf defined by the resistive term (i.e. $- \eta \bf {J}$).  The current
${\bf J}$ used in this step is computed from the partially updated
field resulting from the MOCCT step.  Resistive heating [i.e. the last
term on the LHS of eq.\ (4)] is computed with this same
current, appropriately averaged to the grid center.  Since our update
of the resistive term is time explicit, we also add a new timestep
constraint so that $\bigtriangleup t \leq [\min (\bigtriangleup x,
\bigtriangleup y, \bigtriangleup z)]^{2}/\eta$.

We tested our implementation of the resistivity algorithm by following
the diffusion of a magnetic field with an initially gaussian profile in
a non-rotating box, a problem whose solution is known analytically.  As
described in \S3, we also have reproduced the linear stability
criterion for the MRI in resistive differentially rotating flows.

\subsection{Initial Conditions}

For these simulations we use a computational volume with radial
dimension $L_x$ = 1, azimuthal dimension $L_y = 2\pi$, and vertical
dimension $L_z$ = 1.  Most of the runs use a standard grid resolution
of $59 \times 123 \times 59$.  (Note that this resolution is comparable
to the high-resolution runs of HGB and HGB2.  The increase in what
constitutes a standard grid resolution simply reflects the increase in
computational power over the last few years.) Initially the
computational domain is filled with a uniform plasma of density
$\rho_{0}=1$ and pressure $P_{0}=10^{-6}$.  We set $\Omega = 10^{-3}$,
sound speed $c_{s} = (\gamma P/\rho)^{1/2}$
and vertical scale height $H = c_{s}/\Omega \approx 1.3$.
We study the evolution of a variety of initial magnetic field
configurations including constant vertical fields $B_z$, constant
toroidal fields $B_y$, and spatially varying $B_z$ fields whose
volume-average sums to zero (``zero-net field'').  The initial magnetic
field strength $B_{0}$ is specified by $\beta =
P_{0}/(B_{0}^{2}/8\pi)$.

Each of these initial field configurations is evolved using a range of
resistivities $\eta$.  The importance of a specific value of
resistivity is characterized by the magnetic Reynolds number $Re_{M}$,
defined as a characteristic length times velocity divided by $\eta$.
Here we define $Re_{M}$ in terms of important disk length and velocity
scales, namely, the disk sound speed and vertical scale height,
\begin{equation}
 Re_{M} \equiv \frac{H c_{s}}{\eta} .
\label{reynolds}
\end{equation}
With this definition, the magnetic Reynolds number is independent of
the initial magnetic field.  An alternative definition uses the
wavelength of the fastest growing mode of the MRI ($\sim v_{A}/\Omega$)
and $v_{A}$ as the characteristic length and velocity, 
\begin{equation}
Re_{M}' = \frac{v_{A}^2}{\Omega \eta}.
\label{sano}
\end{equation}  
This definition was used by Sano et al.\ (1998).  The
two definitions are related by $Re_M = Re_M' \beta / 2$ using the
parameter $\beta = 2c_{s}^{2}/v_{A}^{2}$.

The essential properties of resistive MRI can be understood from simple
physical scalings.  In the nonresistive limit the MRI's fastest growing
wavenumber has $k\cdot v_A \approx \Omega$.  For a given resistivity
$\eta$ the magnetic field diffusion rate will be of order $k^2 \eta$.
The MRI will be strongly affected at wavenumbers where the resistive
damping rate exceeds the MRI linear growth rate.  An important
demarcation point is established by setting the resistive damping
wavenumber $k_D$, defined to be that wavenumber where the diffusion
rate is equal to $\Omega$, equal to the wavenumber of the fastest
growing MRI mode, $k_{MRI} = \Omega/v_{A}$.  This occurs when $Re_{M}'
= 1$, or
\begin{equation}
Re_{M} = \beta/2.
\end{equation}

In addition to the critical Reynolds number, established by $k_D =
k_{MRI}$, one can define other important limits.
Because the resistive damping rate is proportional
to the square of the wavenumber, the largest wavenumbers (shortest
wavelengths) of the MRI
will be affected first.  Small wavenumbers
have the potential to remain unstable for larger resistivities.  In
the small wavenumber (large wavelength) limit, the growth rate of the
MRI is proportional to $k\cdot v_{A}$.  If we equate this growth rate
to the resistive damping rate we obtain the condition $v_{A}/k\eta
=1$.  Thus, for a given scale $H$, it is possible to
damp all modes with $k \ge 2\pi/H$, and completely suppress the MRI, if 
\begin{equation}
Re_{M} = 2\pi{c_s\over v_A} \propto \beta^{1/2}.
\end{equation}
In a numerical simulation, the largest available scale is set by the
dimensions of the computational domain, $L$, and this stability
limit would also be proportional to the ratio $H/L$.  In any case,
when the diffusion wavenumber is
$k_D = 2\pi/L$ the entire computational box would be dominated by
diffusion on a timescale $\Omega$.  This Reynolds number is
\begin{equation}
Re_{M} = \left( {2\pi \over L}\right)^2 \left({H c_s \over
\Omega}\right) = (2\pi H/L)^2 \sim 40 (H/L)^2.
\end{equation}

At the other extreme we can also define a Reynolds number for which the
diffusion wavenumber is $k_D = 2\pi/\Delta x$, where $\Delta x$ is the
size of a grid zone.  For a computational grid size $L$ divided into
$N$ grid zones the gridscale Reynolds number is
\begin{equation}
Re_{M} = (2\pi)^2 (HN/L)^2.
\end{equation}
For Reynolds numbers larger than this, diffusion will not be the
dominant effect on dynamical timescales for all computationally
resolved lengthscales.  The magnetic Reynolds numbers used for our
investigations below are all smaller than this limit.

Our numerical simulations also possess an intrinsic numerical resistivity 
due to truncation error.  Because the numerical resistiviy is a
nonlinear function of the grid spacing, it must be measured for each
individual application by increasing the magnitude of the physical resistivity
from zero and noting at what point the solution diverges from the ideal 
case.   As described in section 3.2, we find for our standard resolution our
numerical Reynolds number is about 50,000 for vertical fields with zero net
flux.   

\section{Results}

\subsection{Uniform Vertical Fields}

We begin with simulations of an initial weak uniform poloidal magnetic
field in the shearing box.  Due to periodic boundary conditions, the net
flux associated with this initial mean field remains unchanged for all times.  Hawley \& Balbus
(1992) simulated this problem in 2D in the ideal MHD limit and found
that the instability produced an exponentially growing ``channel
solution'' consisting of two oppositely directed radial streams
surrounded by radial magnetic field.  In 2D, saturation of the
instability does not occur; instead, simulations terminate when the
plasma channel is squeezed into a thin sheet too small to resolve.  In
3D, however, the channels break down into MHD turbulence due to
nonaxisymmetric ``parasitic'' instabilities (Goodman \& Xu 1994; HGB;
HGB2) so long as the vertical wavelength of the channel mode is less
than the radial size of the computational domain.

Even in 2D, however, additional possibilities are created by the
addition of a finite resistivity.  SIM
carried out 2D simulations of the vertical field problem with a
resistive plasma for a variety of field strengths, and for
resistivities $0.3 \leq Re_{M}' \leq 3$ [using the definition of
Reynolds number given by eq.\ (7)].  They found an interesting
dichotomy of behavior depending on whether or not $Re_{M}'$ was greater
or less than one (the critical value).  When $Re_{M}' < 1$, the channel
solution saturates via magnetic diffusion and reconnection
across the radial streams.  Assuming that the resistivity was not so
large as to stabilize all possible wavelengths within the computational
domain, simulations with $Re_{M}' < 1$ amplified the magnetic field
until $Re_{M}' \sim 1$ at which point saturation occurred.  For
stronger initial fields, however, or with weak resistivity such that
$Re_{M}' > 1$ SIM found that the 2D channel solution remained as
before, growing without apparent limit.  Since saturation by parasitic
modes is inherently a nonaxisymmetric process, 3D resistive simulations
are required to explore this regime.

In this section we consider a constant vertical magnetic field in an
initially uniform 3D shearing box.  The linear dispersion relation for a
vertical field with resistivity is given by Jin (1996).  Ignoring
buoyancy terms and assuming a Keplerian background and a displacement
${\bf k} = k\hat z$, a simplified linear dispersion relation can be
written
\begin{equation}
\sigma^4 +2\xi \sigma^3 + (2q^2 + \xi^2+1)\sigma^2 +2\xi (q^2+1)\sigma
-3q^2 + q^4 +\xi^2 =0
\label{dispersion}
\end{equation}
where $\sigma$ is a growth rate in units of orbital frequency, $q=k_z
v_{A}/\Omega$, and $\xi = k_z^2 \eta /\Omega$.
Growth rates as a function of $q$ and $\xi$ are plotted in Figure 1.

In a fully conducting plasma, the fastest growing mode of the MRI
has $q=\sqrt{15}/4$; the corresponding wavelength $\lambda_c$ is given by
\begin{equation} 
\lambda_c = 2\pi \sqrt{16/15} |{\bf v_A}|/\Omega.
\end{equation} 
For these uniform vertical field simulations we select $\beta=400$,
$\lambda_c$ = 0.459.  This wavelength extends over 27 grid cells so
fast-growing modes are well resolved.  The largest vertical wavelength
allowed in the box is $\lambda = L$ which corresponds to $q=0.444$ and
has a growth rate of $\sigma=0.57$.  Wavelengths $L/2$ ($q=0.888$) and
$L/3$ ($q=1.332$) are also unstable with growth rates $\sigma = 0.75$
and 0.66 respectively.  The zero resistivity run serves as a control
model; it uses the same parameters (although a slightly different
numerical resolution) as the run labeled Z4 in HGB.  We have run four
resistive simulations with this initial magnetic field using
$Re_M=1300$, 520, 260, and 130.  For increasing values of magnetic
resistivity the most unstable wavelength shifts to larger scales and
the growth rate is substantially lower than the ideal MHD rate.  The
linear growth rates for the three largest wavelengths with $Re_M =1300$
are $\sigma = 0.53$, 0.62, and 0.41.  From the dimensional analysis  of
the critical magnetic Reynolds number, for our given field strength and
box size, we expect that the growth rate of the MRI should be
significantly affected for $Re_M \sim \beta = 400$.  Indeed, at that
level $\lambda = L/3$ is no longer unstable, and the growth rates for
$\lambda = L$ and $L/2$ are reduced to 0.45 and 0.36.  At $Re_M = 260$
only the $\lambda=L$ wavelength remains unstable with a growth rate of
0.37.  This wavelength too becomes marginally stable for $Re_M = 130$.

Figure 2 is a plot of the time evolution of the volume averaged total
magnetic energy in each run.  In keeping with the linear analysis, we
find no growth when $Re_M=130$; all modes have been stabilized by
resistivity.  Unstable modes are present for all other values of
$Re_M$, with the most rapid growth for the non-resistive
($\eta=0$) model.  For each of the models in which unstable modes
are present, the evolution of the magnetic energy is similar.  An
initial phase of exponential growth ends in a sharp peak, after which
the magnetic energy declines rapidly before fluctuating about a lower
average level.  For $\eta=0$, this initial peak occurs at 3.5
orbits.  Consistent with the decrease in linear growth rates, the peak
occurs at increasingly later times for decreasing $Re_M$; for
$Re_M=260$ the magnetic energy peaks at 8 orbits.  

In the absence of resistivity, the initial linear growth is associated
with development of the channel solution at the wavelength of the
fastest growing mode; for $\eta=0$ this is $L_z/2$.  This is also
the fastest growing wavelength in the $Re_M=1300$ model, which  behaves
similarly.  The growth rate for the $Re_M=1300$ run is $80\%$ of the
maximum growth rate for ideal MHD (0.75$\Omega$), close to what is
predicted by the linear analysis.  The $\lambda = L/2$ channel solution
is well established by orbit 2.2 for $\eta=0$ and 3.2 for
$Re_M=1300$.  A spectral analysis at this stage indicates that the
magnetic energy of the fastest growing wavenumber mode is 3 orders of
magnitude greater than all other modes combined.

In the runs with $Re_M = 520$ and 260, the linear growth rates have
been reduced by magnetic diffusion enough to change the nature of the observed channel
solution.  Ohmic dissipation stabilizes small scale perturbations, and
we no longer observe the $\lambda = L_z/2$ mode; the dominant mode is
now $\lambda = L$, i.e., a two stream (one in, one out) mode.  With
$Re_M=260$ the growth rate of this mode has decreased to $48\%$ of the
ideal MHD value.

As resistivity is increased, the peak in the magnetic energy at the end
of the linear growth phase achieves higher amplitudes.  This suggests
that the fastest growing modes of the parasitic instabilities (the
Kelvin-Helmholtz modes) are also affected by resistivity (Goodman \& Xu
1994).  One reason for this is that
the parasitic modes require vertical wavenumbers that
are larger than radial wavenumbers.  For $Re_M$ less than 260,
resistivity restricts growth to the smallest vertical wavenumber,
inhibiting the onset of the parasitic instability.  This allows the
channel solution to persist and to reach higher amplitudes before
disruption.

Following the peak and subsequent decline, the magnetic energy in all
models fluctuates around a mean value.  Increasing the resistivity (1)
reduces this mean saturated field energy, (2) increases the magnitude
of the fluctuations relative to the saturation amplitude, and (3)
increases the timescale of the fluctuations.  Table 1 lists
volume-averaged values for a variety of quantities in the saturated
state, time-averaged over the last twenty orbits of each run.  There is
a systematic decline in both kinetic and magnetic energies with
increasing resistivity.  For the $Re_M=260$ model the mean kinetic
energy in the turbulent fluctuations is $26\%$ of the mean value in the
ideal MHD simulation.  In the turbulent flow the magnetic energy
resides primarily in the azimuthal component of the field.  The
background shear flow always favors the growth of the azimuthal field
component.  The ratio of $B_y^2$ to $B_x^2+B_z^2$ is 2.6 in the zero
resistivity run, increases to 3.1 for the $Re_M = 520$ run before
dropping back to 2.6 with $Re_M = 260$.  The magnetic field energy is
greater than the perturbed kinetic energy by a factor of 2.4 with no
resistivity; this ratio decreases with increasing resistivity, down to
1.24 in the $Re_M = 260$ run.

Turbulent shear stress $W_{xy} = {\rho v_x\delta v_y - B_xB_y/4\pi}$
transports angular momentum outward, but the total stress decreases
with Reynolds number.  Using Shakura-Sunyaev scaling, $W_{xy}=\alpha
P_o$, $\alpha$ decreases from 0.307 at $\eta=0$ to 0.053 at
$Re_M=260$.  Angular momentum transport is correspondingly reduced.
In all cases the Maxwell stress dominates over the Reynolds stress,
although the ratio of Maxwell to Reynolds stress decreases from 4.8 for
$\eta=0$ to 3.38 for $Re_M=260$.

Although the time histories of these runs (Figure 2) are superficially
similar, the late-time state of the $Re_M=520$ and 260 simulations is
strikingly different than that of the higher $Re_M$ runs.  In the ideal
MHD case, and in the $Re_M=1300$ run, the flows are turbulent.  At
lower $Re_M$ values, however, the fluctuations seen in the saturated
state are associated with the periodic re-emergence of the channel
solution.  Figure 3 shows images of the angular momentum excess $\delta
L \equiv \rho (v_{y} + 1.5\Omega x)$ overlaid by magnetic field lines
in the $x-z$ plane at $y=0$ at orbits 23 and 25 in the ideal
($\eta=0$) and $Re_M=260$ models.  Note from Figure 2 that a
strong peak in the magnetic energy begins to develop in the $Re_M=260$
model at these times.  In the ideal model, the angular momentum excess
shows large amplitude, disordered fluctuations; the magnetic field is
highly tangled and time-variable, characteristic of MHD turbulence.  In
contrast, the $Re_M=260$ model shows a disordered field and pattern of
angular momentum excess at orbit 23, but by orbit 25 (during the growth
of the next peak in the magnetic energy) this has changed to a layered
profile, combined with a large amplitude sinusoidal variation in the
field, both characteristic of the channel solution.  A spectral
analysis of the magnetic field in the $Re_M=260$ model at orbit 25
shows not the power law usually associated with turbulence, but rather
the domination of the $\lambda = L_z$ mode by two orders of magnitude.

The decrease in turbulent flow and the strength of the resurgent
channel solution may be examined by observing the increase in the
kinetic energy associated with radial motions.  In the $Re_M=260$ run, we
find that a large increase in the radial kinetic energy accompanies
every significant increase in magnetic field energy.  For example, at
orbit 23 $<\rho v^2_x> \sim 0.01P_o$ however, by orbit 25 the radial
kinetic energy has increased to $0.1P_o$.  The radial kinetic energy
rises by at least one order of magnitude every time magnetic field
energy peaks.

For $Re_M=260$, the channel solution first re-emerges (after the
initial peak) at orbit 17, and thereafter occurs roughly every 4
orbits.  These recurrent channels saturate not through parasitic
instabilities, but by reconnection across the channels.  In their 2D
simulations, SIM found that reconnection could be an alternative
saturation mechanism for the channel solution; we find that this
behavior in 3D as well.  The resistive diffusion time for variations of
lengthscale $L_z/2$ (the vertical extent of one channel) is
$L_z^{2}/4\eta$, or about 6 orbits.  The frequency of variability
is comparable to this resistive diffusion time.

The quasi-periodic emergence of the channel solution could have
important consequences for time variability of the accretion rate in
disks.  Figure 4 plots the shear stress normalized by the initial
pressure ($W_{xy}/P_{0} = \alpha$) for the ideal and $Re_M$=260 case.
Fluctuations in the stress exceed an order of magnitude, although the
mean is less than the ideal MHD case.  Note that the stress associated
with the initial peak results in $\alpha > 1$.  We may conclude that
for high values of magnetic diffusion in a mean field, the rate of
transport is cyclic and highly variable.

Considerable heating due to Ohmic dissipation occurs,
particularly in the high resistivity
runs.  For example, at $Re_M=260$ the internal energy rises by a factor
of $\sim 100$ during the initial decline from the magnetic energy peak
at orbit 8.  This is a much greater amount of heating than that seen in
the other runs.  It would appear that most of the magnetic energy
extracted from the differential rotation by the growth of the MRI now
heats the disk through Ohmic dissipation.  This heating is an
indication that resistivity is playing a major role in terminating the
channel flow.  During the later phase of evolution, the internal energy
rises in a somewhat stair-step fashion during each re-emergence of the
channel solution, although never again at so large a rate.  For
example, after saturation the internal energy in the $Re_M=260$
simulation rises only another 5\% by orbit 40.  Again this indicates
saturation through resistivity.  Ohmic heating also drives changes in
the plasma $\beta$ parameter.  Since we have a simple adiabatic
equation of state, with no radiative losses, heat simply accumulates.
Initially the MRI drives $\beta$ toward unity.  However, after
saturation $\beta$ differs considerably for runs with different
resistivity, for example $\eta=0$ $\beta\approx 17$ while for
$Re_m=260$ $\beta \approx 3000$.  In these simulations this heating
keeps the magnetic pressure from becoming dynamically important, but has
limited importance since there is no vertical gravity.  In a disk,
radiative cooling would limit the growth of $\beta$.

\subsection{Vertical Fields with Zero Net Flux}

In the previous section we considered a domain filled with a uniform
vertical field.  Since the currents generating such a field are outside
the computational domain, the background flux cannot change,
regardless of the strength of the resistivity.  The next set of
simulations begins with an initial magnetic field configuration $B_z =
B_o \sin 2{\pi}x$.  With periodic (and shearing-periodic) boundary
conditions, the mean field $\langle B\rangle$ remains zero for all
time.  Therefore, unlike the uniform vertical field simulations above,
physical resistivity can now completely dissipate the field.  This
particular initial zero-net field configuration was previously studied
in the ideal MHD limit by HGB2.

We describe the results of six runs corresponding to six values of
magnetic Reynolds number:  $Re_M=\infty(\eta=0)$, 65K, 26K, 19.5K, 13K, and
1.3K.  These simulations are all performed with a maximum field
strength of $\beta=400$ and at our standard resolution.  Figure 5 is a
plot of the time-evolution of the total magnetic energy for these
values of the magnetic Reynolds number.  Except for the case of
$Re_M=1.3$K, an initial period of exponential growth is observed,
followed by saturation in a sharp peak near 3.2 orbits.  As in the
uniform vertical field runs, this rise to a strong  peak is associated
with the $k_x=k_y=0$ and $k_z=4\pi$ mode, the fastest growing mode from
the linear analysis.  Note that in all except the ideal MHD case, some
diffusion of the initial field can be observed during the first two
orbits; for $Re_M$=13K and $Re_M$=1.3K the field energy is $89\%$ and
$24\%$ of its initial value by orbit 2.2.  The diffusion is so rapid
for the $Re_M=1.3$K model that significant growth in the field energy
is never attained.  The azimuthal and radial components of the magnetic
energy peak at $\sim 10^{-5}P_o$, only $\sim$ 4 orders of magnitude
greater than their values at orbit 0.5.  After orbit 4 the energy in
both of these components starts to decline.  The vertical component
does not exhibit any growth at all.  Unlike the radial and azimuthal
field components any amplification is insignificant compared to the
resistive loss.  The growth of the linear mode is, in turn, severely
restricted when the background field is rapidly diffusing.

As $Re_M$ is decreased, there is a small decline in the peak values of
azimuthal and radial field energy; it is the vertical field energy
that is most affected.  At $Re_M$=13K the peak
value of vertical field energy is only 1.24 times greater then its
initial value.  Resistivity affects the evolution not so much by
altering the growth rates of the most unstable modes, but by diffusing
away the background field upon which those modes are growing.  This was
not a factor in the uniform initial field models, and the decline in
the amplitude of the nonlinear flow at saturation as a function of
$Re_M$ is a consequence.

After the initial peak, the magnetic energies decline to an approximate
mean value about which there are large amplitude, short-timescale
fluctuations.  Because the spatial variation of the field in the
initial state causes the wavelength of the fastest growing mode to
vary, parasitic instabilities are not required to cause the channel
solution to transition to turbulence.  Instead, the structure which
results from the growth of the linear modes is already complex enough
that turbulence is the inevitable outcome.  After the initial
peak, the MRI saturates as MHD turbulence for all values of $Re_M$ 
except $Re_M = 1300$.

Table 2 lists the average value in the turbulent state of selected
quantities for several runs.  Each quantity has been averaged over the
time period 30 -- 50 orbits.  For $\eta=0$,  significant growth of
the MRI occurs after 2 orbits, with the saturated magnetic energy level
$\sim 0.01P_o$.  At 2.2 orbits, $95\%$ of the field energy is located
in the toroidal field.  The magnetic energy is dominated by the
azimuthal field with $85\%$ of this energy in the azimuthal component.
The $Re_M$=130K and $Re_M$=65K simulations show little apparent
difference in energy levels.  Angular momentum transport was similar in
these three simulations as well, with $\alpha \sim 0.008$.  Since there
is little difference between resistive runs with $Re_M \gapprox 65$K
and ideal MHD runs, the effective numerical resistivity would appear to
be of the same order, $\sim$ $Re_M=65K$.

Below $Re_M =65$K, there is a systematic decline in the saturation
amplitude with decreasing $Re_M$.  The magnetic energy is $3 \times
10^{-4}P_0$ for $Re_M=13$K, compared with $10^{-2}P_o$ for the ideal
MHD run.  Angular momentum transport also declines with decreasing
Reynolds number.  In the $Re_M=26$K run the field saturates at levels
above the initial energy, with a small but non-negligible transport
rate of $\alpha \gapprox 2 \times 10^{-3}$.  The $Re_M = 13K$ run is
strongly affected.  At orbit 3 it has $\alpha$=0.007 and a Maxwell to
Reynolds stress ratio = 2.4.   The poloidal field is preferentially
destroyed as time proceeds; at orbit 3 $53\%$ of the field energy is in
the $B_y$ component but this increases to $92\%$ by orbit 10.  At orbit
30, $\alpha$ = 0.0001 and $99\%$ of the remaining energy is in the
$B_y$ component.  Transport has effectively been shut down in the disk,
and Ohmic dissipation allows the field energy to continue decreasing.
Although the magnetic energy stops declining beyond orbit 30, the mean
magnetic energy and effective $\alpha$ are so small as to imply the MRI
is effectively quenched.  At this point the field has large scale
organized structure.  It is layered:  in the upper half of the box
field is directed toward the positive azimuthal direction,  and in the
lower half the field is directed toward the negative azimuthal
direction.

We expect the dissipational lengthscale to grow as $Re_M$ is lowered.
This expectation can be examined quantitatively.  Figure 6 is a comparison
of the power spectrum of fluctuations in the magnetic energy as a function
of wavenumber in the y-dimension $k_y$ in the $Re_M=13K$ and the $\eta=0$
runs.  The spectra are time averaged over orbits 15 to 23 for $Re_M=13K$ and
16 to 27 for $\eta=0$.  The $Re_M=13K$ spectrum has been normalized to
give it the same amplitude as the $\eta=0$ case.  Both spectra are fit
by decreasing power laws, with fluctuations on large scales (small k) fit by
a Kolmogorov-like slope (-11/3).  On small scales (large k), where dissipation 
becomes important, the slope of the spectra becomes much steeper.  For the
$Re_M=13K$ run, this change in slope occurs at about $k_y \sim 8 (2\pi)/L_y$,
while for the ideal run the change does not occur until $k_y \sim 11 (2\pi)/L_y$.
Thus, the large explicit resistivity clearly smooths the turbulence on small 
scales as expected. 

In the ideal MHD simulations of HGB2 using an initial vertical field
that varies as $\sin (x)$, the energies in the late-time turbulent
state were more or less independent of the initial field strength.
Here we consider a zero net vertical flux model with $\beta=1600$
initially and $Re_M=19.5K$, computed at the standard resolution.  The
evolution of the magnetic energy is qualitatively similar to the
$\beta=400$ model except that the energy levels of the weaker field
simulation always remain a factor four times lower throughout the
evolution.  The field energy is concentrated in the azimuthal component
for both runs.  Both runs show similar time behavior with respect to
angular momentum transport.  An average from orbit 4 through orbit 10
yields $\alpha=0.012$ for the $\beta=400$ model; this is five times
greater then average transport for the $\beta=1600$ run.  The power
spectra for these runs are also similar.

One difference between the $\beta=400$ and the weak field $\beta=1600$
simulation shows up clearly in the linear stage.  The weak field run
has a critical wavelength $\lambda_c=L _z/4$.  This is consistent with
linear analysis, which, for $\beta$=1600 predicts a critical wavelength
$\lambda_c=0.23$.  Decreasing the field strength while keeping $Re_{M}$
and the box size fixed simply shifts the wavelength of the most
unstable mode down closer to the resistive dissipation scale.  Since
the growth rate of the MRI and the resistive dissipation rate are
unchanged, the subsequent time evolution of volume averaged variables
is similar to the $\beta=400$ run, although the energies remain lower
and the detailed structures of the nonlinear flow are different.

\subsection{Toroidal Fields}

In this section we consider the effect of resistivity on the
development of the MRI in the presence of a toroidal field.  The linear
properties of the MRI in the local ideal MHD limit were considered by
Balbus \& Hawley (1992).  The toroidal field instability is
nonaxisymmetric, and, for a nonaxisymmetric mode, the radial wavenumber
$k_R$ evolves with time due to the background shear.  For a pure toroidal
field, amplification occurs only during that time when $k/k_z$ is
small.  Although peak amplification still occurs for wavenumbers
$k\cdot v_A \approx\Omega$ ($k$ here corresponds to the azimuthal
wavenumber for a toroidal field), the toroidal field instability favors
large values of $k_z$ ($\rightarrow\infty$) and these are the
wavenumbers most likely to be affected by resistivity.  The vertical
field instability, on the other hand, favors finite $k_z$ and
$k_R=k_\phi = 0$ (axisymmetric channel solutions).

Papaloizou and Terquem (1997)  examined in some detail the linear
stability of a toroidal field configuration with finite resistivity.
Consistent with the expectations from the nonresistive linear analysis,
they found that mode growth ceased for magnetic Reynolds numbers ($\sim
1000$) that are larger than those that stabilize the vertical field
instability.  Interestingly, their condition for transient
amplification, their equation (32), corresponds to the zero-frequency
limit of the linear dispersion relation for the poloidal field
instability.  Again many of the qualitative linear properties of the
MRI are essentially independent of field strength or orientation.

We have run several simulations of shearing boxes with initial constant
toroidal fields.  Such simulations were carried out in the nonresistive
limit by HGB who examined a variety of initial field strengths.  Here
we choose an initial field strength of $\beta =100$.  Models were run
at both $64\times 128\times 64$ and $32\times 64\times 32$ grid
resolution using initial random perturbations at about 1\% of the
sound speed.  We will concentrate on the high resolution simulations
which were computed for the ideal MHD limit and for $Re_{M} = 10K$, 5K
and 2K.

The evolution of the toroidal ($B_y$) and radial ($B_x$) magnetic
energies is shown in Figure 7.  As the Reynolds number is reduced from
the ideal MHD limit, the initial growth rates are reduced and final
saturation is delayed.  The $Re_M = 2K$ model shows no growth.
Post-saturation time-averaged values for the simulations are given in
Table 3.  All values show a decline with decreasing Reynolds number,
but there is a sharp transition from perturbation growth to
perturbation decay in going from $Re_M=5K$ to $2K$.  This transition is
consistent with the results for certain specific linear modes carried
out by Papaloizou \& Terquem (1997).

\subsection{Resolution}

As a step toward investigating the effect of finite resolution on our
simulation results, we performed two resolution experiments.  We reran
the case of vertical field with zero net flux at four different
resolutions, once with zero resistivity, and once for $Re_M = 10$K.
The resolutions ran from $16\times 32\times 16$ up to $128\times
256\times 128$ by powers of two.  The initial magnetic field is the
same as in \S3.2 above, a vertical field that varies as $\sin(x)$.  A
specific set of long-wavelength initial velocity perturbations is
applied to each model so that all resolutions begin with the same
initial conditions.  

The resolution series with zero resistivity behaves in an expected
way.  The higher the resolution, the larger the rate of growth of the
perturbed magnetic field energy, the higher the initial peak, and the
earlier that peak occurs.  Beyond the initial peak the magnetic energy
declines somewhat.  The lowest resolution model exhibits a noticeably
larger rate of decline, while the other three resolutions look
comparable.

Figure 8 plots the time evolution of the magnetic energy in the series
of runs with $Re_M=10$K.  The behavior of the different resolution
models is the same as in the zero resistivity run for the first few
orbits.  After this the magnetic energy declines for all resolutions;
the highest resolution run, however, shows the steepest rate of decline
in magnetic energy, followed by the $64\times 128\times 64$ resolution
run.  The lowest resolution runs decline less steeply and behave
similarly to each other.  For this particular Reynolds number, the
critical diffusion wavenumber, defined $\eta k^2 = \Omega$,
corresponds to a wavelength of 0.0714 (where the vertical box size $L_z
= 1$).  This wavelength is equal to $1.14\Delta z$ in the lowest
resolution simulation (and hence is unresolved), and equal to
$18.3\Delta z$ in the highest resolution simulation.  Thus we have the
interesting observation that by resolving the diffusion lengthscales,
turbulence decays more rapidly in the highest resolution grid.
Apparently ``numerical resistivity'' is much less effective at field
dissipation compared to a physical resistivity with a diffusive
wavelength comparable to the grid zone size.

\section{Summary}

Using numerical MHD simulations, we have studied the shearing box
evolution of the MRI in the presence of finite resistivity.  We have
examined initial field configurations consisting of a uniform vertical
field, a uniform toroidal field, and a vertical field that varies
sinusoidal in the radial direction.  The linear growth rates of the
most unstable modes observed in our simulations are in good agreement
with a linear analysis.   As the resistivity increases, the growth rate
for all modes declines.  For a fixed box size, all modes are stable at
a large enough value of the resistivity (when $\eta \gapprox
v_{A}^{2}/\Omega$); our numerical results correctly recover this
limit.  The restrictions imposed upon the toroidal field instability
are more severe.  We find toroidal field models become linearly stable
at larger Reynolds numbers than vertical field models, again, in
agreement with the linear analysis (Papaloizou \& Terquem 1997).
Because the toroidal field instability favors large vertical
wavenumbers, the modes that are the most unstable and have the longest
period of amplification are precisely those most affected by finite
resistivity.

Although the simulations agree with the linear analysis
during the linear growth phase, we find that the nonlinear evolution
is more complicated, and the linear analyses provide  only limited
guidance.  In particular, finite resistivity can have profound effects
on the flow even when the linear modes are still unstable.

The nonlinear outcome of the MRI is profoundly influenced by the
presence or absence of a net field.
When the shearing box is penetrated by a net vertical field, Ohmic
dissipation can never completely destroy the mean field.  Thus,
provided the resistivity is low enough that at least a few unstable
modes are present, the MRI will always exist in such a box.  When
the resistivity is large, but not so large as to completely stabilize
the MRI,  the instability leads to a strongly fluctuating magnetic
energy and transport rate.  These fluctuations are associated with
periodic recurrence of the axisymmetric channel solution.  High
resistivity can then lead to reconnection across channels and rapid
decline in magnetic energy, after which the instability grows again.
The  period of the resurgent channel solution is found to be roughly
equal to the resistive diffusion time, of order a few orbits.

It should be stressed that saturation through reconnection is observed
only when the initial field is supported by some external currents
outside our simulation domain.  Also, the effects of stratification are
not present in our simulations.  These considerations may limit the
degree to which our results may be generalized to realistic disk
models.  However, locally in highly resistive disks threaded by a mean
magnetic field transport may be cyclic.  This result may be of some
importance for our understanding time variable accretion systems.

We find that models with a non-vanishing net magnetic flux display
qualitatively different behavior in the nonlinear regime than models
with zero net flux.  If the initial field configuration contains  zero
net magnetic flux then the resistive MRI saturates as MHD turbulence,
as it does in the ideal MHD case.  Finite resistivity, however,
significantly modifies the evolution of the turbulence.  The amplitude
of the magnetic energies and corresponding angular momentum transport
rates in these simulations decline with decreasing $Re_M$.  In fact,
for $Re_M \lapprox 10^4$ the turbulence is completely quenched.  This
limit is roughly 100 times larger than the Reynolds number required
for complete stabilization within the linear
theory.  Examination of the power spectrum of the turbulence clearly
shows a rapid drop off in power at at high wavenumbers (small scales)
when resistivity is present compared to the ideal MHD limit.

The finding that finite resistivity can affect the levels of turbulence
even when the linear analysis predicts the presence of instability, has
potential implications for accretion disk evolution.  In particular,
Gammie \& Menou (1998) point out that finite resistivity leads to
magnetic Reynolds numbers $Re_M \le 10^4$ in the cool, low states of
dwarf novae.  Our simulations show that this is indeed an interesting
level of resistivity.  Many dwarf nova models depend upon different
levels of angular momentum transport in the high and low state.  Finite
resistivity appears to be a viable mechanism by which these different
levels could be produced.

\acknowledgements
This research is supported in part by NASA grant NAG-54278 and NSF
grant AST-9528299 (JMS), and NASA grants NAG5-7500 and NAG5-3058, and
NSF grant AST-9423187 (JFH).  We thank Steve Balbus and Charles Gammie
for insightful comments.  Supercomputer simulations are supported under
an NSF National Resource Allocation grant, and have been carried out on
the Origin 2000 system at NCSA, and the T90 and T3E systems at NPACI.

\newpage

\begin{figure}
\plotone_reduction{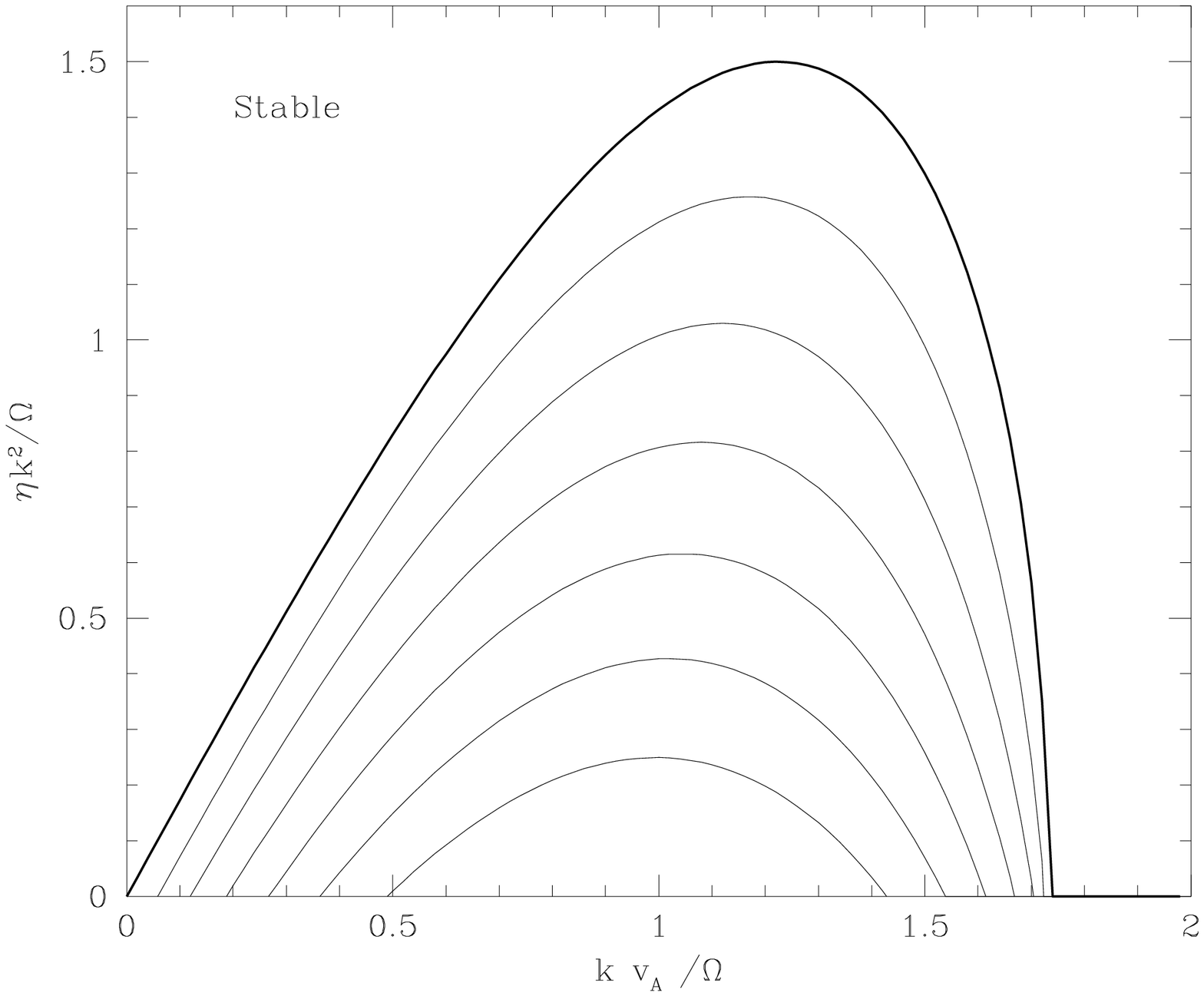}{0.9}
\caption{Linear growth rates for vertical field instability as a
function of resistive frequency $\eta k^2/\Omega$ versus MRI frequency
$k\cdot v_A/\Omega$.  From top to bottom the curves equal growth rates
of 0 (stability boundary), 0.1, 0.2, 0.3, 0.4, 0.5 and 0.6 $\Omega$.}
\end{figure}

\begin{figure}
\plotone_reduction{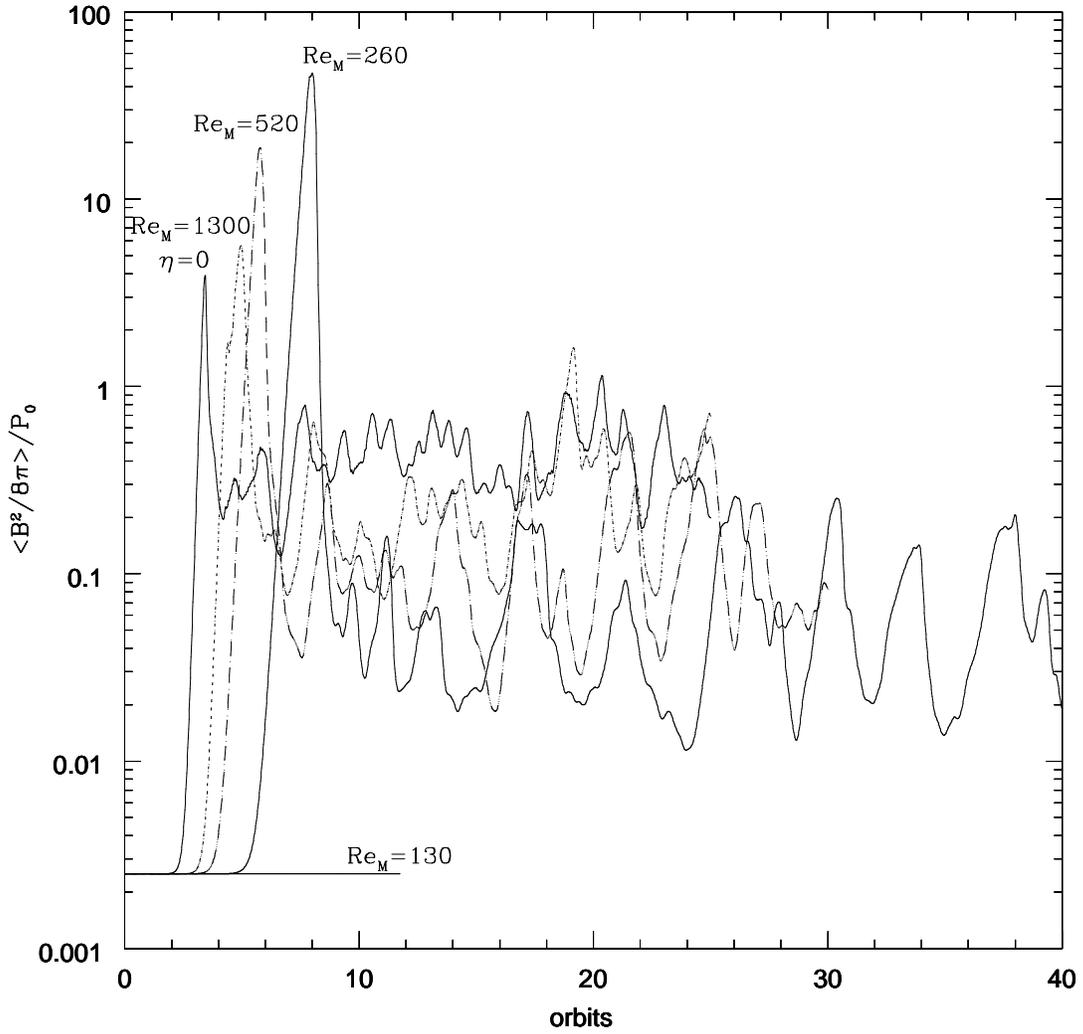}{0.9}
\caption{Time evolution of the volume averaged magnetic energy
in simulations beginning with a uniform vertical field with $\beta=400$
and various $Re_M$.}  
\end{figure}

\begin{figure}
\plotone_reduction{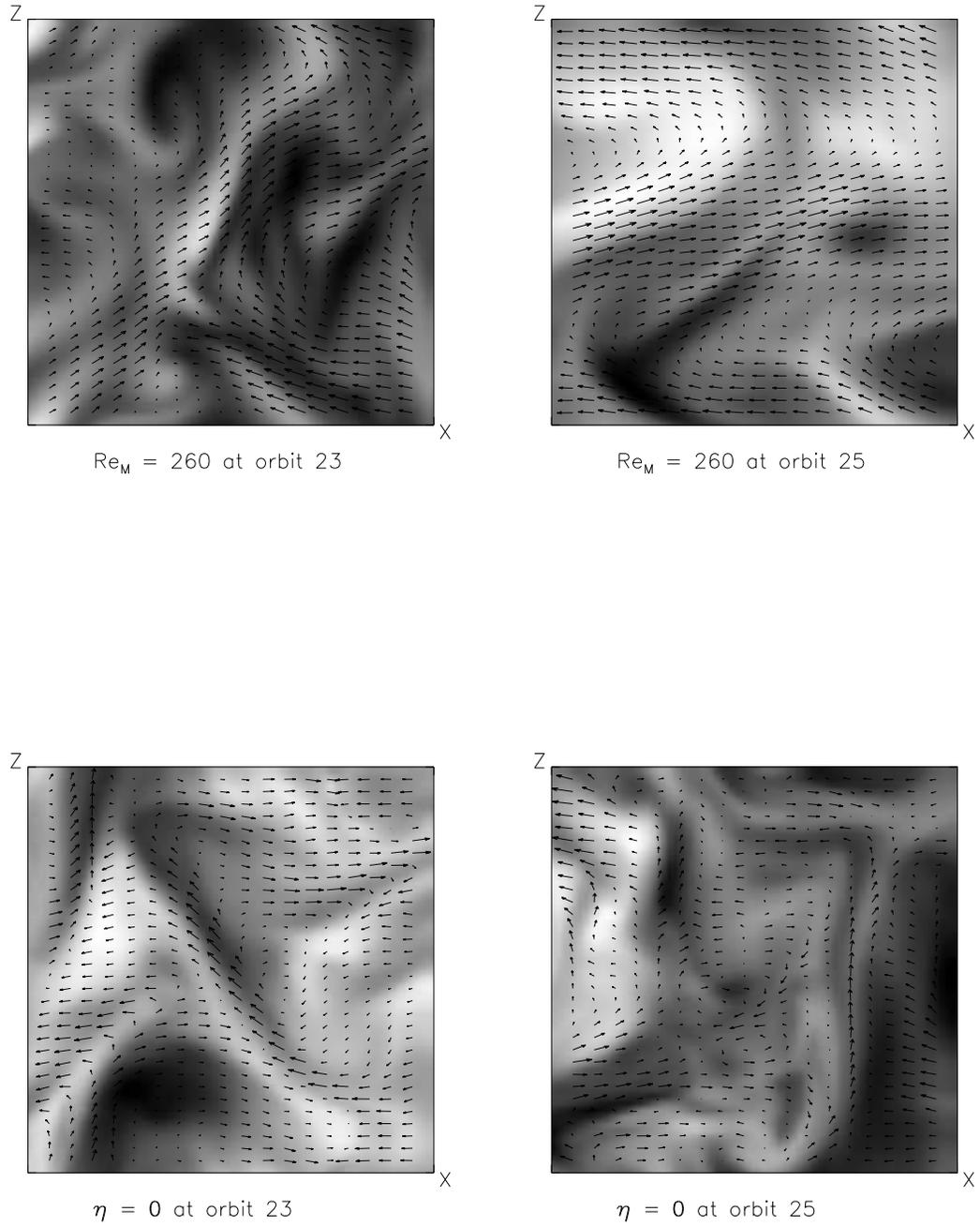}{0.9}
\caption{Slices in the $x-z$ plane at $y=0$ of the angular
momentum excess (colors) and poloidal magnetic field (arrows)
at orbits 24 and 26 in the $Re_M=260$ and $\eta=0$ runs.
Note the re-emergence of the channel solution in the resistive case.}
\end{figure}

\begin{figure}
\plotone_reduction{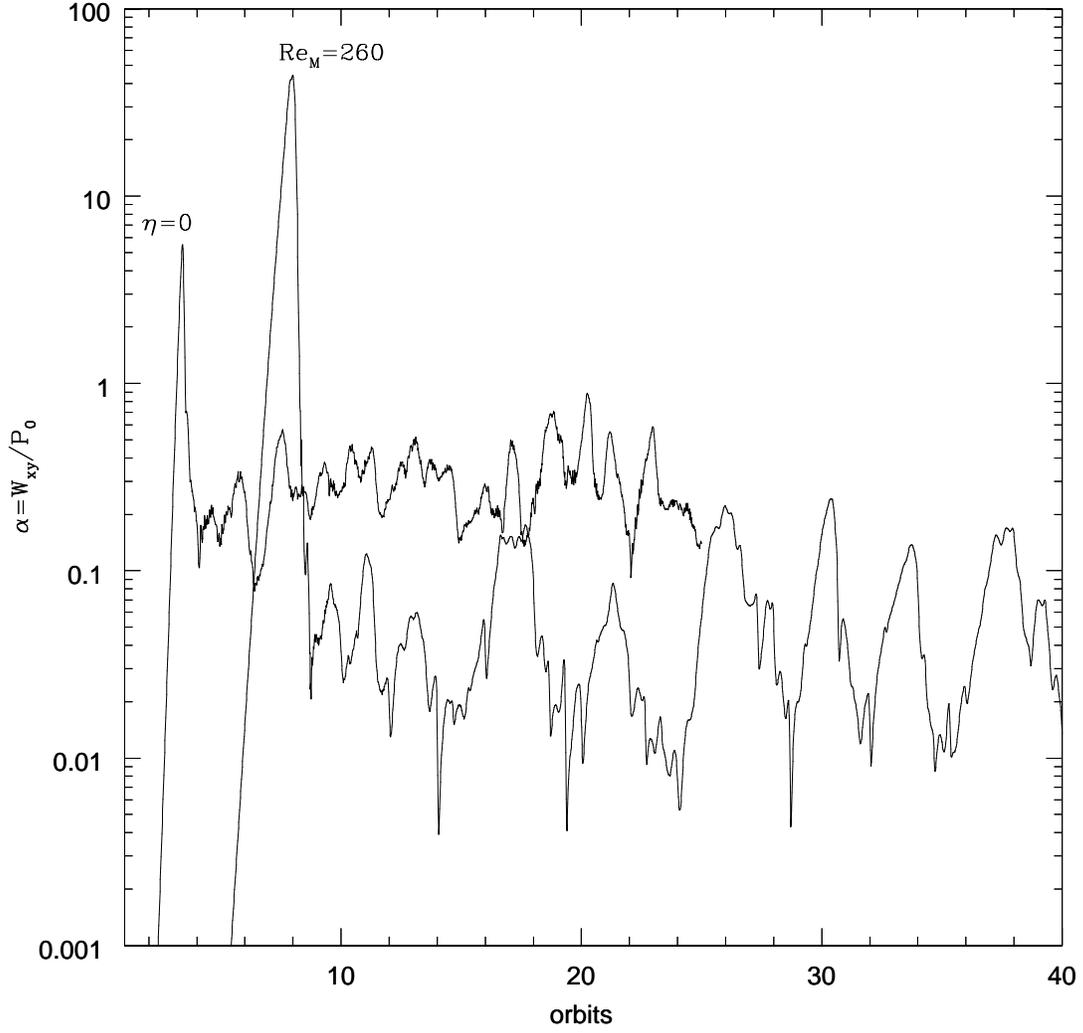}{0.9}
\caption{Evolution of $\alpha = W_{R\phi}/P_{0}$ in uniform vertical field
models.  There are large fluctuations in the angular momentum transport
rate in the resistive case.}
\end{figure}

\begin{figure}
\plotone_reduction{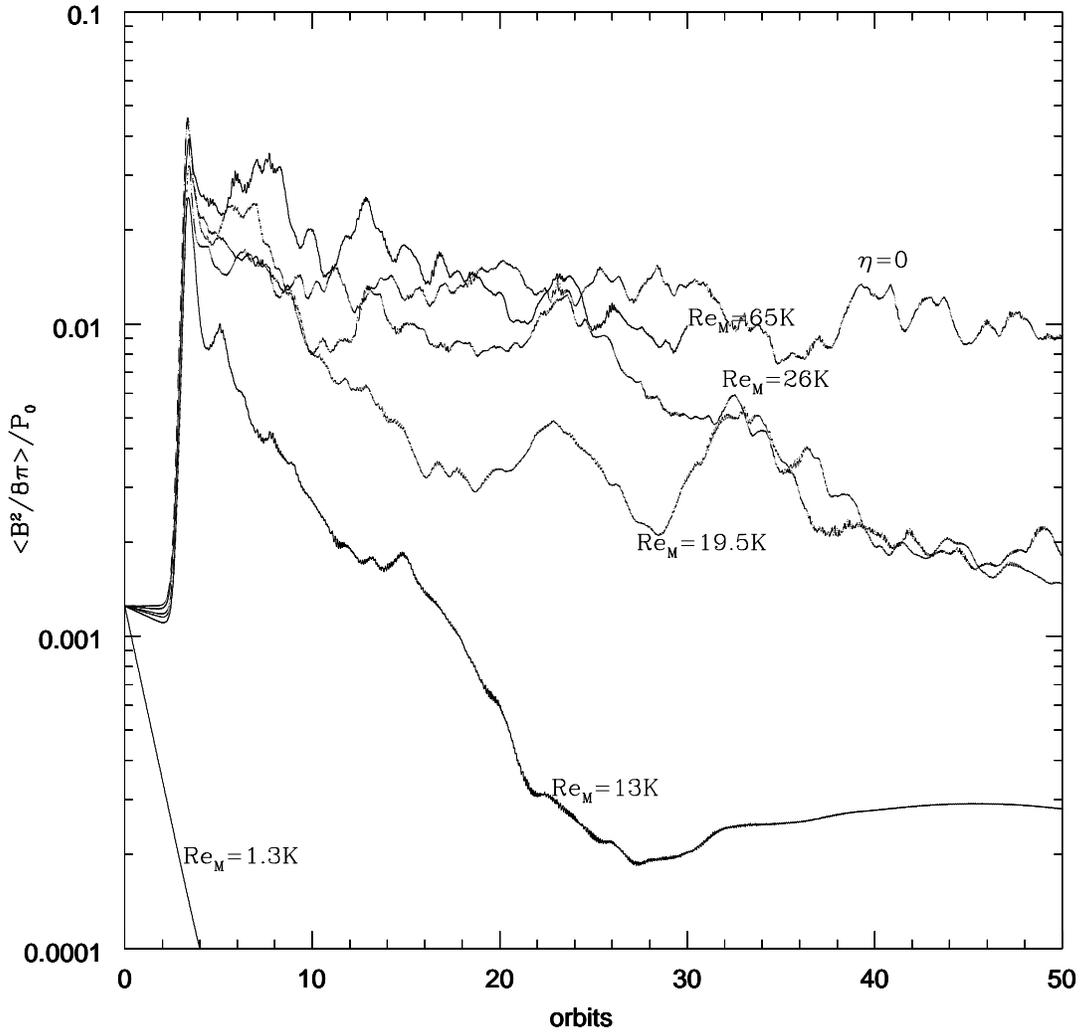}{0.9}
\caption{Time evolution of the volume averaged magnetic energy
in simulations beginning with a zero net flux vertical field with $\beta=400$
and various $Re_M$.}
\end{figure}

\begin{figure}
\plotone_reduction{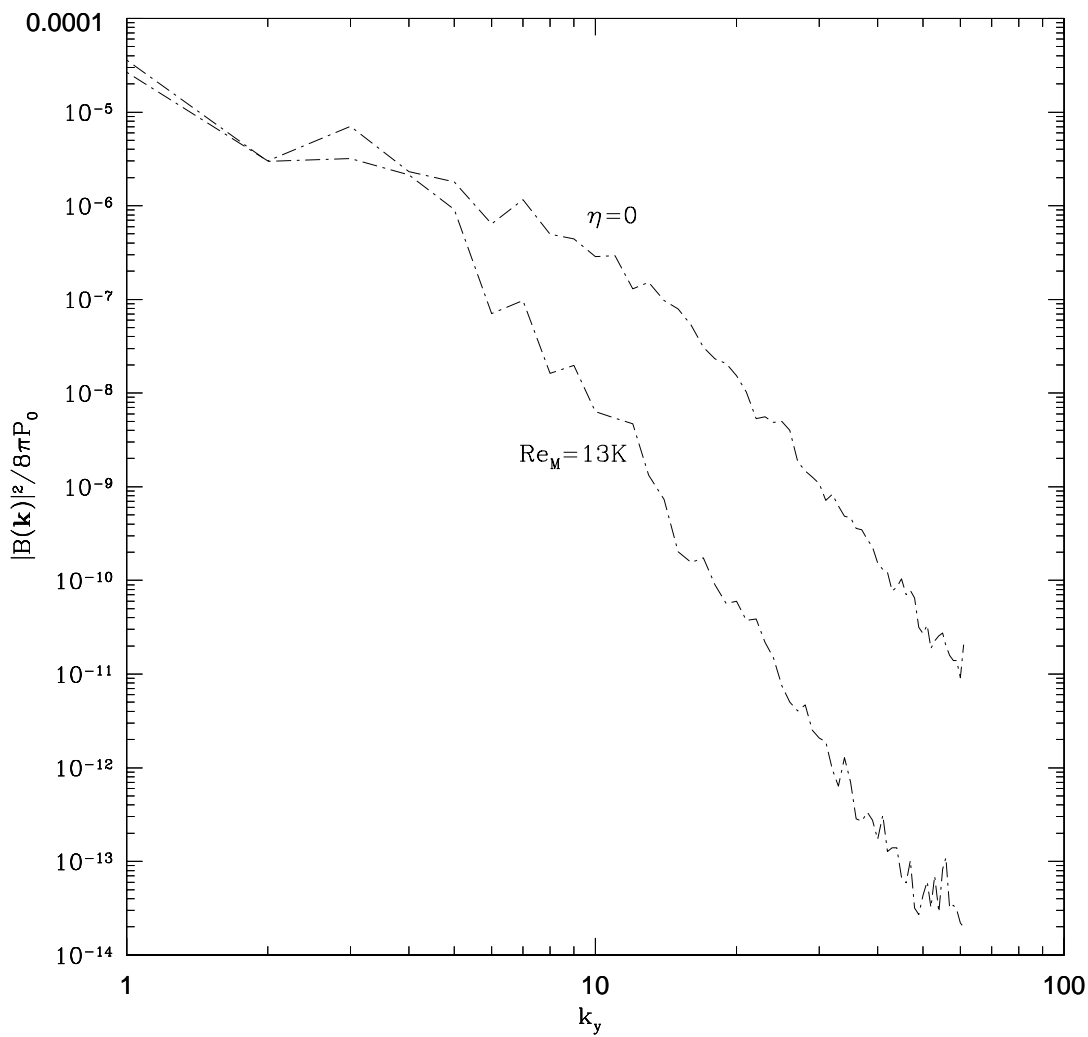}{0.9}
\caption{Comparison of the power spectrum of fluctuations in the magnetic energy as 
a function of $k_y$.  Here $\eta=0$ and $Re_M=13K$ are displayed in units of
$2\pi/L_y$.  The $Re_M=13K$ run has been normalized to give it the same amplitude
as the $\eta=0$ simulation.  Both spectra are fit by a Kolmogorov-like slope
(-11/3) on large scales.}
\end{figure}

\begin{figure}
\plotone_reduction{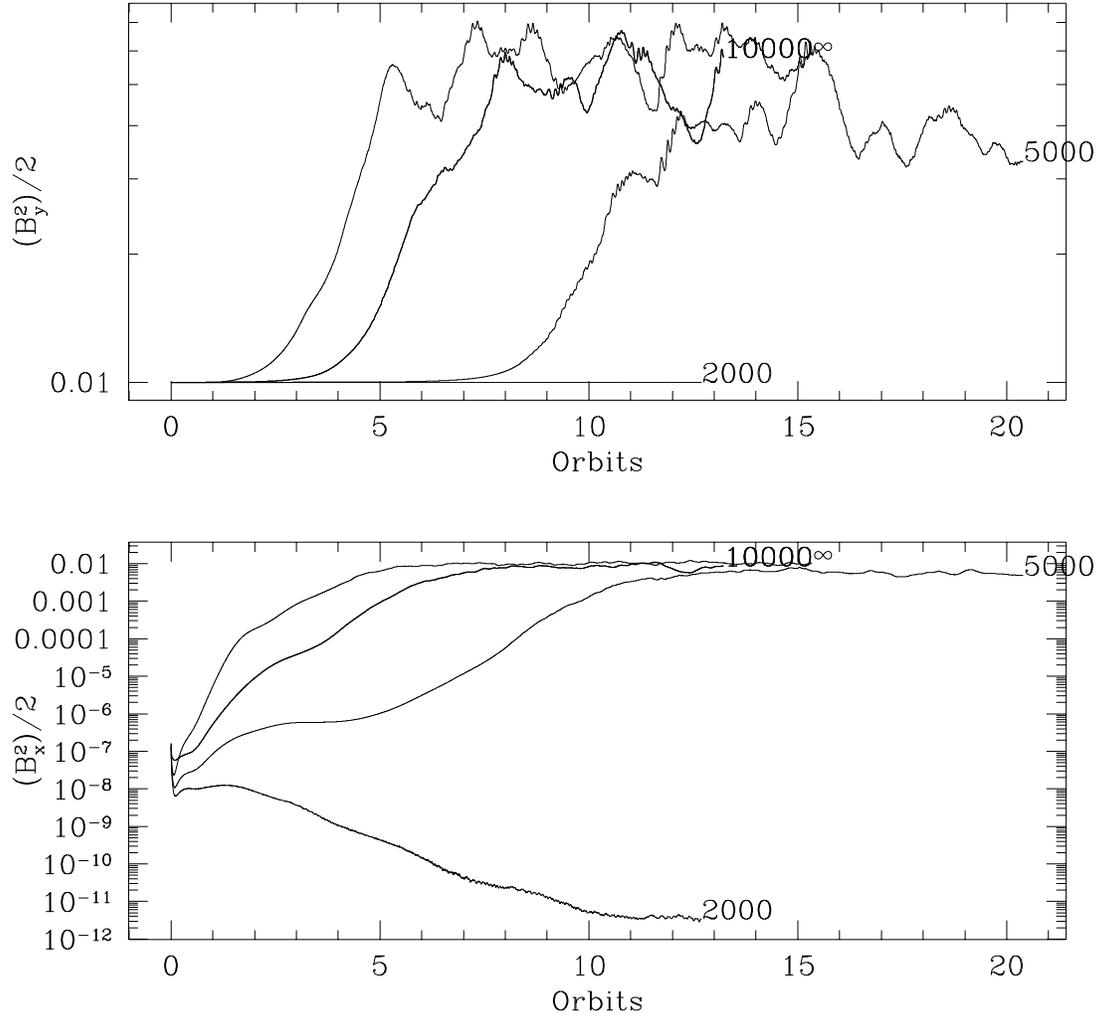}{0.9}
\caption{Time evolution of the toroidal and radial magnetic energies for simulations
beginning with a uniform toroidal field for varius $Re_M$.}
\end{figure}

\begin{figure}
\plotone_reduction{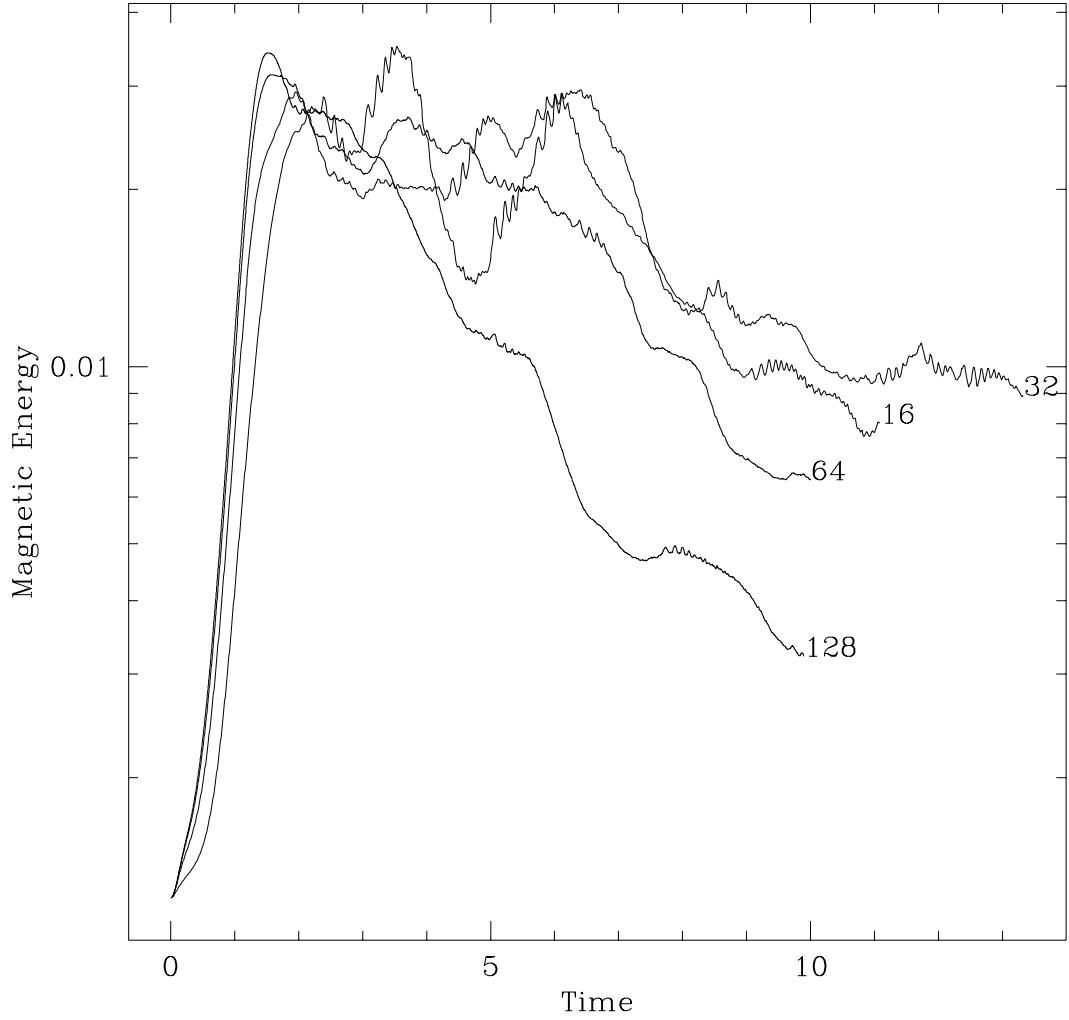}{0.9}
\caption{Time evolution of the volume averaged magnetic energy in simulations 
beginning with a zero net flux vertical field at $Re_M=10K$ for various resolutions.}
\end{figure}

\clearpage

\newpage

\begin{center}
{\bf TABLE 1} \\ TIME- AND VOLUME- AVERAGE VALUES FOR UNIFORM VERTICAL FIELD RUNS \\
            
\addvspace {0.5cm}

\begin{tabular}{ccccc} \hline \hline
\\ [-0.3cm]
Quantity & 
$\eta=0$ &
$Re_M=1300$ & 
$Re_M=520$ & 
$Re_M=260$ \\  
\hline
\\ [-0.2cm]

$B^2/8\pi P_o$ & 0.450 & 0.277 & 0.140 & 0.062 \\  
$B^2_x/8\pi P_o$ & 0.093 & 0.055 & 0.025 & 0.012 \\
$B^2_y/8\pi P_o$ & 0.324 & 0.204 & 0.105 & 0.045 \\
$B^2_z/8\pi P_o$ & 0.033 & 0.018 & 0.009 & 0.005 \\
$-B_xB_y/4\pi P_o$ & 0.254 & 0.170 & 0.087 & 0.042 \\
$\rho v_x \delta v_y/P_o$ & 0.052 & 0.040 & 0.022 & 0.012 \\
$\rho \delta v^2/2P_o$ & 0.192 & 0.135 & 0.081 & 0.050 \\
$\rho v^2_x/2P_o$ & 0.071 & 0.059 & 0.040 & 0.028 \\
$\rho \delta v^2_y/2P_o$ & 0.092 & 0.052 & 0.024 & 0.011 \\
$\rho v^2_z/2P_o$ & 0.029 & 0.024 & 0.017 & 0.011 \\
$\alpha$ & 0.307 & 0.210 & 0.110 & 0.053 \\
Max/Reyn & 4.8 & 4.23 & 3.87 & 3.38 \\
\hline
\end{tabular}
\end{center}

\newpage
\begin{center}
{\bf TABLE 2} \\ TIME- AND VOLUME- AVERAGE VALUES FOR ZERO MEAN VERTICAL FIELD RUNS \\
           
\addvspace {0.5cm}

\begin{tabular}{ccccc} \hline \hline
\\ [-0.3cm]
Quantity & 
$\eta=0$ & 
$Re_M=26K$ &
$Re_M=19.5K$ &
$Re_M=13K$  \\  
\hline
\\ [-0.2cm]

$B^2/8\pi P_o$ & 0.010 & 0.003 & 0.003 & 0.00027  \\
$B^2_x/8\pi P_o$ & 0.001 & 0.00016 & 0.00014 & $6.10 \times 10^{-8}$  \\
$B^2_y/8\pi P_o$ & 0.0089 & 0.003 & 0.0026 & 0.00027  \\
$B^2_z/8\pi P_o$ & 0.0003 & $5.38 \times 10^{-5}$ & $4.0 \times 10^{-5}$ & $8.28 \times 10^{-9}$ \\
$-B_xB_y/4\pi P_o$ & 0.0045 & 0.001 & 0.0009 & $3.75 \times 10^{-6}$ \\
$\rho v_x \delta v_y/P_o$ & 0.0023 & 0.001 & 0.00069 & $2.28 \times 10^{-5}$  \\
$\rho \delta v^2/2P_o$ & 0.008 & 0.004 & 0.0025 & 0.0001 \\
$\rho v^2_x/2P_o$ & 0.0051 & 0.003 & 0.002 & 0.0001 \\
$\rho \delta v^2_y/2P_o$ & 0.002 & 0.00075 & 0.00042 & $2.57 \times 10^{-5}$ \\
$\rho v^2_z/2P_o$ & 0.001 & 0.0003 & 0.0002 & $8.94 \times 10^{-7}$ \\
$\alpha$ & 0.0068 & 0.0011 & 0.0016 & $2.65 \times 10^{-5}$ \\
Max/Reyn & 1.92 & 0.870 & 1.29 & 0.165 \\
\hline
\end{tabular}
\end{center}

\newpage
\begin{center}
{\bf TABLE 3} \\ TIME- AND VOLUME- AVERAGE VALUES FOR TOROIDAL FIELD RUNS \\
           
\addvspace {0.5cm}

\begin{tabular}{ccccc} \hline \hline
\\ [-0.3cm]
Quantity & 
$\eta=0$ & 
$Re_M=10K$ &
$Re_M=5K$ &
$Re_M=2K$  \\  
\hline
\\ [-0.2cm]

$B^2/8\pi P_o$ & 0.0705 & 0.061 & 0.049 & ---  \\
$B^2_x/8\pi P_o$ & 0.0099 & 0.0081 & 0.0059 & ---  \\
$B^2_y/8\pi P_o$ & 0.0568 & 0.050 & 0.0403 & 0.01  \\
$B^2_z/8\pi P_o$ & 0.0038 & $0.0031$ & $0.0024 $ &---\\
$-B_xB_y/4\pi P_o$ & 0.030 & 0.026 & 0.0102 & --- \\
$\rho v_x \delta v_y/P_o$ & 0.0077 & 0.0036 & 0.00069 & ---  \\
$\rho \delta v^2/2P_o$ & 0.0304 & 0.026 & 0.021 & ---  \\
$\rho v^2_x/2P_o$ & 0.0138 & 0.012 & 0.0097 & --- \\
$\rho \delta v^2_y/2P_o$ & 0.0113& 0.0090 & 0.0071 & --- \\
$\rho v^2_z/2P_o$ & 0.0053 & 0.0048 & 0.0045 & --- \\
$\alpha$ & 0.039  & 0.034  & 0.014 & --- \\
Max/Reyn & 3.33 & 3.38  & 2.83 & --- \\
\hline
\end{tabular}
\end{center}

\end{document}